\UseRawInputEncoding

\documentclass[12pt]{article}
\usepackage{longtable,supertabular}
\usepackage{amsmath}
\usepackage{amssymb}
\usepackage{latexsym}
\usepackage{cite,a4wide,color}
\usepackage{multirow}
\usepackage{ntheorem}
\usepackage[square,numbers]{natbib}

\theoremstyle{remark}

\title{\bf Griesmer and Optimal Linear Codes from the Affine Solomon-Stiffler Construction}
\author{Hao Chen \thanks{Hao Chen is with the College of Information Science and Technology/Cyber Security, Jinan University, Guangzhou, Guangdong Province, 510632, China, haochen@jnu.edu.cn. This research was supported by NSFC Grant 62032009.
}}

\begin{document}
	
	\maketitle
	\begin{abstract}
		In their fundamental paper published in 1965, G. Solomon and J. J. Stiffler invented infinite families of codes meeting the Griesmer bound. These codes are then called Solomon-Stiffler codes and have motivated various constructions of codes meeting or close the Griesmer bound. However weight distributions of Solomon-Stiffler codes have been only determined for very special cases.\\

In this paper, we give a geometric construction of affine and modified affine Solomon-Stiffler codes. Projective Solomon-Stiffler codes are special cases of our modified affine Solomon-Stiffler codes. Several infinite families of $q$-ary Griesmer, optimal, almost optimal two-weight, three-weight, four-weight and five-weight linear codes are constructed as special cases of our construction.  Weight distributions of these Griesmer, optimal or almost optimal codes are determined explicitly. Many optimal linear codes documented in Grassl's list are re-constructed as (modified) affine Solomon-Stiffler codes.\\

Several infinite families of  optimal or Griesmer codes were constructed in two published papers in IEEE Transactions on Information Theory 2017 and 2019, via Gray images of codes over finite rings. Parameters and weight distributions of these Griesmer or optimal codes and very special case codes in our construction are the same. We also indicate that more general distance-optimal binary linear codes than that constructed in a recent paper of IEEE Transactions on Information Theory can be obtained directly from codimension one subcodes in binary Solomon-Stiffler codes.\\

		 Modified affine Solomon-Stiffler code. Weight distribution.
	\end{abstract}
	
	\newpage

\section{Introduction}

\subsection{Preliminaries}

We recall some basic facts in the coding theory, see \cite{HP}. The Hamming weight $wt({\bf a})$ of a vector ${\bf a}=(a_0, \ldots, a_{n-1}) \in {\bf F}_q^n$ is the cardinality of nonzero coordinates. The Hamming distance $d({\bf a}, {\bf b})$ between two vectors ${\bf a}$ and ${\bf b}$ is $d({\bf a}, {\bf b})=wt({\bf a}-{\bf b})$. Then ${\bf F}_q^n$ is a finite Hamming metric space. The minimum (Hamming) distance of a code ${\bf C} \subset {\bf F}_q^n$ is, $$d({\bf C})=\min_{{\bf a} \neq {\bf b}} \{d({\bf a}, {\bf b}),  {\bf a} \in {\bf C}, {\bf b} \in {\bf C} \}.$$  One of the main goals in the theory of error-correcting codes is to construct codes ${\bf C} \subset {\bf F}_q^n$ with large cardinalities and minimum distances. In general, there are some upper bounds on cardinalities or minimum distances of codes.  Optimal codes attaining these bounds are particularly interesting, see \cite{HP}. A linear code over the finite field ${\bf F}_q$ with the length $n$, dimension $k$ and the minimum Hamming weight $d$ is denoted as a linear $[n,k,d]_q$ code. A linear code is called projective if its dual distance is at least three. Let $A_i({\bf C})$, $i=0,1,\ldots, n$, be the number weight $i$ codewords in the linear code ${\bf C} \subset {\bf F}_q^n$. Then it is obvious that there are at most $n-d({\bf C})+1$ nonzero weights $i \in \{d({\bf C}), d({\bf C})+1, \ldots, n\}$, such that $A_i({\bf C}) \neq 0$. If $A_i({\bf C}) \neq 0$ for exact $t$ indices $i_1<i_2<\cdots<i_t$, this linear code is called $t$-weight linear code. A simplex code is an one-weight code.  When $t$ is small, constructions of few-weight linear codes and the determination of their exact weight distributions are important topic in coding theory.\\

For fixed $q,n,k$, if there is a linear $[n,k,d]_q$ code, and there is no $[n,k,d+1]_q$ code. We call this code optimal. If there is a linear $[n,k,d+1]_q$ code, but there is no linear $[n, k, d+2]_q$ code, we call this linear $[n,k,d]_q$ code almost optimal. If there is a linear $[n,k,d+2]_q$ code, but there is no linear $[n, k, d+3]_q$ code, we call this $[n,k,d]_q$ code near optimal.\\

\subsection{Related works}

\subsubsection{Griesmer codes and Solomon-Stiffler codes}

The Griesmer bound for a linear $[n, k, d]_q$ code in \cite{Griesmer} asserts $$n \geq \Sigma_{i=0}^{k-1} \lceil \frac{d}{q^i} \rceil.$$ Set $g_q(k,d)=\Sigma_{i=0}^{k-1} \lceil \frac{d}{q^i} \rceil$, for a linear $[n,k,d]_q$ code, the Griesmer defect is $g({\bf C})=n-g_q(k,d)$. A linear $[n, k, d]_q$ code attaining this bound, that is, the Griesmer defect zero code, is called a Griesmer code, see \cite{Solomon,Hu1}. In 1965, Solomon and Stiffler constructed infinite families of binary and $q$-ary Griesmer codes in \cite{Solomon}. In 1973, Baumert and McEliece proved that for fixed $q$ and $k$, there are Griesmer codes for any given fixed large minimum distance $d$, see \cite{BM}. Then many authors constructed Griesmer codes from various methods, see \cite{Hell,Hell2,Hell1,Hell3,BJV,Maruta,Liu,DingTang,Hu,Sihem3,Hu1,DingHeng}. Optimal linear codes close to the Griesmer bound were also constructed in \cite{Shi,Hu,Hu1,Sihem2}. However, codes in \cite{KM,Maruta} have Griesmer defects one or two.\\

In coding theory, it is extremely difficult to determine exact weight distributions of linear codes. For constructions of few-weight linear codes from Boolean functions and designs, and the determination of their exact weight distributions, we refer to \cite{carlet,Ding,DD1,DingHeng,Hell4,Sihem2, Sihem1,DingTang,Hu,Hu1,Xu} and references therein.\\

We recall the definition of binary Solomon-Stiffler code associated with $h$ integers $u_1, \ldots, u_h$, satisfying $$k>u_h>u_{h-1}>\cdots>u_1\geq 1.$$ Suppose that there are $h$ dimension $u_1, \ldots, u_h$ subspaces $S_1, \ldots, S_h$ in ${\bf F}_2^k$, such that $S_i \bigcap S_j ={\bf 0}$. Let $F$ be the set of disjoint nonzero column vectors in dimension $u_1$, $u_2$, \ldots, $u_h$ subspaces $S_1, \cdots, S_h$ in ${\bf F}_2^k$. Deleting these columns in the generator matrix of the binary simplex $[2^k-1,k,2^{k-1}]_2$ code, we obtain a binary linear $[2^k-1-\Sigma_{i=1}^h (2^{u_i}-1),k, 2^{k-1}-\Sigma_{i=1}^h 2^{u_i-1}]_2$ code meeting the Griesmer bound, see \cite{Solomon}. Similarly, $q$-ary projective Solomon-Stiffler codes can be defined, we refer to \cite{Solomon}. In particular, when $h=1$, $u_1=k-1$, 1st order Reed-Muller code is a binary Solomon-Stiffler code associated with one positive integer $u_1=k-1$. However, weight distributions of general Solomon-Stiffler codes have not been determined completely. Binary Solomon-Stiffler codes were also used to construct optimal binary linear complementary pairs in \cite{Guneri}.\\

In \cite[Theorems 3, 5]{Hell3}, it was showed that projective binary or $q$-ary Griesmer codes are projective Solomon-Stiffler codes, when their minimum weights satisfy some condition. In the case of a binary Griesmer $[n,k,d]_2$ code, if $d \leq 2^{k-1}$, this code is binary Solomon-Stiffler or Belov code. For $q$-ary Griesmer codes, we refer conditions on $d$ to \cite[Theorem 5]{Hell3}. Notice that Griesmer codes constructed in this paper are not projective. Then these Griesmer codes are not $q$-ary projective Solomon-Stiffler codes determined by Theorem 3 and Theorem 5 of \cite{Hell3}.\\

\subsubsection{Optimal codes from codes over finite rings}

Several infinite families of $q$-ary linear Griesmer codes or optimal codes have been constructed in \cite{Shi,Shi1,Shi2,Liu,Sihem3} as Gray images of two-weight codes over finite rings.  Weight distributions of these Griesmer and optimal codes were determined in these two papers \cite{Shi,Liu}. In \cite{Shi}, Shi, Guan and Sol\'{e} constructed optimal codes in \cite[Theorem 6.1, 6.3]{Shi} via Gray images of some two-weight codes over the finite ring ${\bf F}_p+u{\bf F}_p$ and determined their exact weight distributions. The main contribution of \cite{Liu} is the construction of Griemser codes in \cite[Theorem 3.5]{Liu}, based on a similar method.  Similarly, Mesnager, Qian and Cao constructed optimal codes in \cite[Theorem 5.4]{Sihem3} via Gray images of codes over a finite chain ring. In \cite{Hu}, more general trace codes, which reach the Griesmer bound or are distance-optimal were constructed. In \cite[Theorm 2]{Shi2} and \cite[Proposition 4.3, 5)]{ML}, some binary $[2^k-2^u,k-1, 2^{k-1}-2^{u-1}]_2$ codes were constructed. In a recent paper  \cite{Vega}, punctured and shortening optimal codes were obtained from Griesmer cyclic codes.\\

\subsubsection{Grassmann codes and related codes}

Projective and affine Grassmann codes and their duals were proposed and studied in \cite{Nogin,BGH1,BGH2} by D. Y. Nogin, P. Beelen, S. Ghorpade and T. H{\o}holdt. Weight spectrum were calculated for some special Grassmann codes, see \cite{KP,PS2}. Fast decoding algorithms of Grassmann codes were studied in \cite{PS1,PS}.\\

Let $m$ be a positive integer and $l <m$ be another positive integer, then linear affine Grassmann $[q^{l(m-l)}, \displaystyle{m \choose l}, q^{l(m-l)}\prod_{i=1}^l(1-\frac{1}{q^i})]_q$ codes $C(l,m)$ were constructed. It is interesting to observe that when $l$ is small, for example, $l=2$ and $m=4$, these Grassmann $[q^4, 6, (q-1)(q^3-q)]_q$ codes have relative small Grisemer defects at most $q^2-q$. Therefore, these affine codes are optimal or close to optimal. For example, when $q=2$, an affine Grassmann $[16,6,6]_2$ code is obtained. This is an optimal binary code, see \cite{Grassl}. When $m=3$, an affine $[81, 6, 48]_3$ code is obtained. The optimal ternary code in this case is a linear $[81, 6, 51]_3$ code, see \cite{Grassl}. On the other hand, affine Solomon-Stiffler $[q^4-q^2, 4, (q-1)(q^3-q)]_q$ code is constructed in Section 2. It seems interesting to find some possible relations between affine Grassmann codes and affine Solomon-Stiffler codes. There have been many papers on codes related to Grassmannians, see \cite{PL} and references therein.

\subsection{Our contributions}

The contributions in this paper are summarized as follows.\\

1) We construct affine Solomon-Stiffler codes and modified affine Solomon-Stiffler codes and upper bound their Griesmer defects in some cases. It shows that many Griesmer, optimal and almost optimal linear codes can be obtained from our construction.\\

2) In Sections 2 and 3, we show that these Griesmer codes and optimal codes constructed in \cite{Shi,Liu} and very special case codes in our construction have the same parameters and weight distributions. Optimal linear $q$-ary codes constructed in \cite{Sihem3} have same parameters as very special codes in our construction. It is also indicated that Griesmer and optimal codes constructed in this paper are less restrictive than those constructed in \cite{Hu}. In Section 7, we construct much more general distance-optimal binary codes than those constructed in \cite{Shi2,ML}. Our distance-optimal codes are codimension one or two subcodes of binary Solomon-Stiffler codes. Distance-optimal punctured codes of Griesmer codes are also discussed in Section 8. Our results in Section 7 and 8 are more general than main results of \cite{Vega}.\\

3) Many optimal linear codes documented in \cite{Grassl} are reconstructed as affine Solomon-Stiffler codes and modified affine Solomon-Stiffler codes, see Table 1 and Table 6. Then these optimal codes have an unified construction and structural. Similarly, many almost optimal and near optimal linear codes are constructed as (modified) affine Solomon-Stiffler codes. This gives a natural geometric structure on these almost optimal and near optimal linear codes.\\

4) Several infinite families of Griesmer or optimal $q$-ary $t$-weight linear codes, $q\leq 4$, $t\leq 5$, are constructed and their weight distributions are determined. Comparing with previous works on few-weight linear codes in \cite{carlet,Ding,DD1,Sihem2, Sihem1,DingTang,Xu}, codes in these infinite families are Griesmer, optimal or almost optimal. Moreover these codes are more general than codes constructed in previous papers, such as, \cite{DD1,Shi,Liu,Shi1,Sihem2,Sihem3}.\\

The paper is organized as follows. In Section 2, affine Solomon-Stiffler codes are introduced. Though these codes are not Griesmer as projective Solomon-Stiffler codes, their Griesmer defects can be upper bounded in some cases. In Section 3, several infinite families of optimal affine Solomon-Stiffler codes are given and their exact weight distributions are determined. In Section 4, modified affine Solomon-Stiffler codes are presented. Comparing to affine Solomon-Stiffler codes, their Griesmer defects decrease. Therefore, modified Solomon-Stiffler codes are more closer to the Griesmer bound. In Section 5, some infinite families of Griesmer modified affine Solomon-Stiffler codes are constructed and their weight distributions are determined. In Section 6, an infinite family of almost optimal codes is given, In Section 7, distance-optimal binary codes are constructed from codimension one subcodes of Griesmer codes. In Section 8, distance-optimal punctured codes of Griesmer codes are give. Section 9 concludes the paper.\\

\section{Affine Solomon-Stiffler codes}

Let $k$ be a positive integer satisfying $k \geq 3$. Let $u_1$, \ldots, $u_h$ be positive integers satisfying $u_1 \leq u_2\leq \cdots\leq u_h$. Let $S_1, \ldots, S_h$ be $h$ linear subspaces of ${\bf F}_q^k$ of dimensions $u_1, \ldots, u_h$ satisfying\\
1) $S_i \bigcap S_j ={\bf 0}$, if $i \neq j$;\\
2) $\Sigma_{i=1}^h(q^{u_i}-1)<q^k-q^{k-1}$.\\

Let $s_1$ be the number of dimension $u_1$ subspaces, $s_2$ be the number of dimension $u_{s_1+1}$ subspaces,..., $s_t$ be the number of largest dimension $u_{s_1+s_2+\cdots+s_{t-1}+1}$ subspaces, where $s_1+s_2+\cdots+s_t \leq h$. Set $g=(s_1-1)+\cdots+(s_t-1)$. {\bf In the case of $q=2$, an important difference to Solomon-Stiffler codes is that the condition $s_i \leq 1$ was imposed in \cite{Solomon}.}\\

Let ${\bf g}_1, \ldots, {\bf g}_{q^k-1}$ be all nonzero vectors of ${\bf F}_q^k$. Let $S$ be the set of all vectors among these $q^k-1$ nonzero vectors in ${\bf F}_q^k$ by deleting these nonzero vectors in $S_1 \bigcup \cdots \bigcup S_h$. Then $$|S|=(q^k-1)-\Sigma_{i=1}^h(q^{u_i}-1).$$

{\bf Lemma 1} {\em $S$ is not contained in any $(k-1)$ dimension subspace of ${\bf F}_q^k$.}\\

{\bf Proof.} Otherwise, we have $|S|\leq q^{k-1}-1$. This is a contradiction to the condition 2).\\

From Lemma 1, the column vectors in $S$ form a $k \times |S|$ matrix $G$ of the rank $k$. Let ${\bf C}$ be the code with the generator matrix $G$.  For the convenience,  we denote the above affine Solomon-Stiffler code by parameters $(k, {\bf u})=(k, u_1, \ldots, u_h)$. Then we have the following result.\\

{\bf Theorem 1} {\em The linear code ${\bf C}$ is a linear $[q^k-1-\Sigma_{i=1}^h (q^{u_i}-1), k, d\geq (q-1)(q^{k-1}-\Sigma_{i=1}^h q^{u_i-1})]_q$ code. This code has at most $2^h$ weights.}\\

{\bf Proof.} For any hyperplane $H \subset {\bf F}_q^k$, it is clear that $|H \bigcap S_i|$ is minimal if and only if $H\bigcap S_i$ is a hyperplane in $S_i$. Then $H-H \bigcap S$ is maximal, when each $H\bigcap S_i$ is a hyperplane in $S_i$, for $i=1, \ldots, h$. The weight of the nonzero codeword corresponding to $H$ is $|S|-|(H-H\bigcap S)|$. The first conclusion follows directly. Since each $|H\bigcap S_i|$ has two possible values, then $|H \bigcap S|$ has at most $2^h$ possible values. The second conclusion is proved.\\

Notice that the condition $s_i \leq q-1$ similar to that in the definition of Solomon-Stiffler codes is not imposed. If $h$ is large and there are many same $u_i$'s. The Griesmer defect is large. The lower bound $$d \geq (q-1)(q^{k-1}-\Sigma_{i=1}^h (q^{u_i-1})$$ is not tight. For example, if $u_1=\cdots=u_h=1$, and $h$ is large, many columns have to be deleted, a hyperplane cannot intersect each of such line at the zero point, see Section 6 below.\\

The lower bound on minimum distances of affine Solomon-Stiffler codes is simple. However, from this simple lower bound and a direct calculation on $$\Sigma_{i=0}^{k-1}\lceil \frac{(q-1)(q^{k-1}-\Sigma_{i=1}^h q^{u_i-1})}{q^i}\rceil,$$ many Griesmer codes and optimal codes can be constructed, as in the following Theorem 2, Corollary 1 and 2.\\

{\bf Lemma 2} {\em Let ${\bf C}$ be a linear $[n,k,d]_q$ code with the Griesmer defect $g$. Then the optimal distance of a linear $[n,k]_q$ code is at most $d+g$.}\\

{\bf Proof.} The conclusion follows from the Griesmer bound immediately.\\

We upper bound Griesmer defects in the following result.\\

{\bf Theorem 2.} {\em The Griesmer defect of this code ${\bf C}$ can be upper bounded in the following several cases.\\
1) If $u_1<u_2<\cdots<u_h$, the code ${\bf C}$ is Griesmer.\\
2) If $u_1=u_2=\cdots=u_h$ and $h \leq q$, the Griesmer defect is upper bounded by $h-1$.\\
3) If $u_1=\cdots=u_{s_1}<u_{s_1+1}=\cdots=u_{s_1+s_2}< \cdots <u_{s_1+\cdots+s_{t-1}+1}=\cdots=u_{s_1+\cdots+s_t}$ and $u_{s_1+\cdots+s_j}\leq u_{s_1+\cdots+s_j+1}-2$ for $j=1,2, \ldots, t-1$. Then the Griesmer defect of the linear code ${\bf C}$ is upper bounded by $g$.}\\

{\bf Proof.} 1) We observe that $\lceil \frac{(q-1)(q^{k-1}-q^{u_1-1}-q^{u_2-1}-\cdots-q^{u_h-1})}{q^{u_1}}\rceil=(q-1)q^{k-1-u_1}-\\ \lfloor \frac{(q-1)(q^{u_1-1}+q^{u_2-1}+\cdots+q^{u_h-1})}{q^{u_1}}\rfloor =(q-1)q^{k-1-u_1}$, since $(q-1)(q^{u_1-1}+\cdots+q^{u_h-1})<q^{u_1}$, from the condition $u_1<u_2<\cdots<u_h$. Then the right side of the Griesmer bound is $\Sigma_{i=0}^{k-1} \lceil\frac{(q-1)(q^{k-1}-q^{u_1-1}-\cdots-q^{u_h-1})}{q^i} \rceil=(q-1)(\frac{q^k-1}{q-1}-\Sigma_{j=1}^h \frac{q^{u_j}-1}{q-1})=(q^k-1)-\Sigma_{j=1}^h(q^{u_j}-1)$.\\

2) We observe that $\lceil \frac{(q-1)(q^{k-1}-hq^{u_1-1})}{q^{u_1}}\rceil=(q-1)q^{k-1-u_1}-\lfloor \frac{h(q-1)(q^{u_1-1})}{q^{u_1}}\rfloor =(q-1)q^{k-1-u_1}-\lfloor \frac{h(q-1)}{q}\rfloor \leq (q-1)q^{k-1-u_1}-h+1$. On the other hand, the term  $\lceil \frac{(q-1)(q^{k-1}-hq^{u_1-1})}{q^{u_1+1}}\rceil=(q-1)q^{k-u_1-2}-\lfloor \frac{h(q-1)}{q^2}\rfloor=(q-1)q^{k-2-u_1}$, from the condition $h\leq q$. Then the conclusion follows directly.\\

3) We calculate the Griesmer defect. Notice that $\lceil \frac{(q-1)(q^{k-1}-\Sigma_{i=s_1+1}^{s_1+s_2}q^{u_i-1})}{q^i}\rceil \\\leq (q-1)\lceil \frac{(q^{k-1}-\Sigma_{i=s_1+1}^{s_1+s_2}q^{u_i-1})}{q^i}\rceil-(s_2-1)$, only when $i=u_{s_1+1}$. On the other hand, if $u_{s_1+s_2+\cdots+s_j+1}\geq u_{s_1+s_2+\cdots+s_j}+2$, this kind of decreasing happens only when $i=u_{s_j+1}$.  Then the conclusion follows.\\

The calculation of the Griesmer defects reveals that, affine Solomon-Stiffler codes introduced in this paper are not Griesmer as Solomon-Stiffler codes in \cite{Solomon}. However their Griesmer defects are small in many cases. Hence, many affine Solomon-Stiffler codes are close to the Griesmer bound, and are distance-optimal or distance-almost optimal, distance-near optimal, with respect to the Griesmer bound, when $h$ is small, see Table 1 below.\\

The following result follows from Theorem 2 directly.\\

{\bf Corollary 1} {\em Let $m$ be a positive integer satisfying $m \geq 3$. Let $(k,{\bf u})=(m+2,1,2,m)$. Then we construct an infinite family of Griesmer $[q^{m+2}-q^m-q^2-q+2, m+2, (q-1)(q^{m+1}-q^{m-1}-q-1)]_q$ codes.}\\

{\bf Proof.} Let $S_2$ be the $2$-dimensional subspace with the base $(1,0,\ldots,0)$ and $(0,1,0, \ldots,0)$ and $S_3$ be the $m$-dimensional linear subspace with the base $(0,0,1, \ldots, 0)$, \ldots, $(0,0,0, \ldots, 1)$. Then $S_1$ is a line spanned by $(1,1,\ldots,1)$. It is obvious that the condition 1) holds. The conclusion then follows from Theorem 2 immediately.\\

{\bf Example 1} Let $q$ be a prime power. When $(k,{\bf u})=(4m+7, m, m+1, m+2, m+3)$, an infinite family of $q$-ary Griesmer linear $$[q^{4m+7}-q^{m+3}-q^{m+2}-q^{m+1}-q^m+3, 4m+7, (q-1)(q^{4m+6}-q^{m+2}-q^{m+1}-q^m-q^{m-1}]_q$$ code are constructed from Theorem 2.\\

Let $q$ be a prime power and $m$ be the positive integer satisfying $m \geq 2$. When $(k,{\bf u})=(4m+3, m, m, m+1, m+2)$, an infinite family of $q$-ary optimal linear $$[q^{4m+3}-q^{m+2}-q^{m+1}-2q^m+3, 4m+3, (q-1)(q^{4m+2}-q^{m+1}-q^{m}-2q^{m-1}]_q$$ code are constructed from Corollary 1.\\

We have the following result for distance-optimal codes.\\

{\bf Corollary 2} {\em Suppose that $u_1 \geq 2$ and only one $s_j$ is $2$ and other $s_i$'s are $1$. Moreover, $$u_{s_1+\cdots+s_{j-2}}\leq u_{s_1+\cdots+s_{j-1}} -2\leq u_{s_1+\cdots+s_{j-1}+1}-4.$$ This linear code ${\bf C}$ constructed in Theorem 1 is distance-optimal with respect to the Griesmer bound.}\\

{\bf Proof.} The conclusion follows from a similar calculation as in the proof of Theorem 2.\\

{\bf Example 2} When $q=2$, $(k, {\bf u})=(5,1,2,3)$, we construct a binary linear Griesmer $[20,5,9]_2$ code, from Theorem 2. This is a binary Solomon-Stiffler code.\\

{\bf Example 3} When $q=7$, $k=2$ and $u_1=u_2=1$, the code in Theorem 1 is a linear $[36,2,30]_7$ code. This code is only almost optimal. The optimal distance is $31$, see \cite{Grassl}. This example illustrates that the condition $u_1 \geq 2$ in Corollary 2 is necessary.\\

{\bf Example 4} When $q=5$, $(k,{\bf u})=(3,2)$, we construct an optimal $[100,3,80]_5$ code. The optimal linear $[100,3,80]_5$ code in \cite{Grassl} was constructed as the shortening code of a linear $[101,4,81]_5$ code, with a speical generator matrix documented in \cite{Grassl}. However, our construction is structural and simpler.\\

Let $q$ be a prime power and $s$ be a positive integer satisfying $s\geq 3$. The affine Solomon-Stiffler code associated with parameters $(2s-2, s)$ is a linear Griesmer $[q^{2s-2}-q^{s-1}, 2(s-1), (q-1)(q^{2s-3}-q^{s-2}]_q$ code. The $q$-copy repetition code is a linear $[q^{2s-1}-q^s, 2(s-1), (q-1)(q^{2s-2}-q^{s-1})]_q$ code. This code is optimal when $s\geq q$ as argued in \cite[Theorem 5.4]{Sihem3}. Two optimal linear codes have the same parameters. The construction of optimal codes in \cite{Sihem3} is based on Griesmer codes over a finite chain ring. Our construction is simpler.\\

In the following table, we give some optimal, almost optimal and near optimal codes constructed in Theorem 2, Corollary 1, and 2, see \cite{Grassl}.

\begin{longtable}{|l|l|l|l|}
\caption{\label{tab:A-q-5-3} Affine Solomon-Stiffler codes}\\ \hline
$q$&$[n,k,d]_q$&$(k,{\bf u})$&Optimality \\ \hline
$3$ &$[52,4,34]_3$& $(4,1,3)$ &optimal\\ \hline
$3$ &$[64,4,42]_3$& $(4,2,2)$ &optimal\\ \hline
$3$ &$[68,4,44]_3$& $(4,1,1,2)$ &almost optimal\\ \hline
$3$ &$[70,4,46]_3$& $(4,1,2)$ &optimal\\ \hline
$3$ &$[74,4,48]_3$& $(4,1,1,1)$ &almost optimal\\ \hline
$3$ &$[78,4,52]_3$& $(4,1)$ &optimal\\ \hline
$3$ &$[206,5,136]_3$& $(5,1,2,3)$ &optimal\\ \hline
$3$ &$[208,5,138]_3$& $(5,2,3)$ &optimal\\ \hline
$3$ &$[214,5,142]_3$& $(5,1,3)$ &optimal\\ \hline
$3$ &$[226,5,5,150]_3$& $(5,2,2)$ &optimal\\ \hline
$3$ &$[232,5,154]_3$& $(5,1,2)$ &optimal\\ \hline
$3$ &$[232,5,152]_3$& $(5,1,1,1,1,1)$ &near optimal\\ \hline
$3$ &$[234,5,154]_3$& $(5,1,1,1,1)$ &near optimal\\ \hline
$3$ &$[236,5,156]_3$& $(5,1,1,1)$ &almost optimal\\ \hline
$3$ &$[238,5,158]_3$& $(5,1,1)$ &almost optimal\\ \hline
$3$ &$[240,5,160]_3$& $(5,1)$ &optimal\\ \hline
$4$ &$[45,3,33]_4$& $(3,1,2)$ &optimal\\ \hline
$4$ &$[48,3,36]_4$& $(3,2)$ &optimal\\ \hline
$4$ &$[54,3,39]_4$& $(3,1,1,1)$ &almost optimal\\ \hline
$4$ &$[57,3,42]_4$& $(3,1,1)$ &almost optimal\\ \hline
$4$ &$[60,3,45]_4$& $(3,1)$ &optimal\\ \hline
$4$ &$[225,4,168]_4$& $(4,2,2)$ &optimal\\ \hline
$4$ &$[237,4,177]_4$& $(4,1,2)$ &optimal\\ \hline
$4$ &$[243,4,180]_4$& $(4,1,1,1,1)$ &near optimal\\ \hline
$4$ &$[246,4,183]_4$& $(4,1,1,1)$ &almost optimal\\ \hline
$4$ &$[249,4,186]_4$& $(4,1,1)$ &almost optimal\\ \hline
$4$ &$[252,4,189]_4$& $(4,1)$ &optimal\\ \hline
$5$ &$[96,3,72]_5$& $(3,1,2)$ &optimal\\ \hline
$5$ &$[116,3,92]_5$& $(3,1,2)$ &almost optimal\\ \hline
$8$&$[49,2,42]_8$& $(2,1,1)$ &almost optimal\\ \hline
$9$ &$[64,2,56]_9$& $(2,1,1)$ &almost optimal\\ \hline
\end{longtable}

Notice all codes in Table 1 are not Solomon-Stiffler codes in \cite{Solomon}. Moreover, it is easy to determine exact weight distributions of optimal, almost optimal and near optimal codes in Table 1, based on a similar geometric method as the proof of Theorem 1.\\

In \cite[Theorem 1]{Hu}, similar codes with some strong restrictions on dimensions of subfields of ${\bf F}_{q^k}$ are studied. Only codes in the case when $u_i$ is not a factor of $u_j$, if $j>i$, or all $u_i$'s are the same, are studied. Comparing with \cite{Hu}, we also need the key condition 1). However it is clear that Theorem 2 gives more general Griesmer and optimal codes than trace Griesmer and optimal codes given in \cite{Hu}. It also should be indicated that distance-optimal codes constructed in \cite[Theorem 9]{Chenhu} have the same parameters as very special codes of our construction. The binary Griesmer codes constructed in \cite[Theorem 7]{Chenhu} are actually classical binary Solomon-Stiffler codes.\\

\section{Optimal few-weight affine Solomon-Stiffler codes}

In this section, we construct several infinite families of optimal two and three weight $q$-ary linear codes, which are based on Theorem 2 and Corollary 1.\\

Let us consider the case $k=2m$, $u_1=u_2=m$. Then we construct a linear $[q^{2m}-2q^m+1, 2m, (q-1)(q^{2m-1}-2q^{m-1})]_q$ code. From Theorem 2, the Griesmer defect is one and this is an optimal code. These codes have the same parameters as optimal codes constructed by Shi, Guan and  Sol\'{e} in \cite[Theorem 6.1]{Shi}. The optimal codes in \cite[Theorem 6.2]{Shi}, are just a repetition copy of the above code. Their $q$ needs to be an odd prime satisfying $q \equiv 3$ $od$ $4$, and $m$ needs to be odd. It should be indicated that our codes are more general. Moreover optimal codes in \cite{Shi} were constructed via Gray images of two-weight codes over a finite ring.\\

{\bf Theorem 3} {\em Let $q \geq 3$ be a prime power. Let $(k,{\bf u})=(2m,m,m)$. Then we construct an infinite family of optimal two-weight $[q^{2m}-2q^m+1,2m, (q-1)(q^{2m-1}-2q^{m-1})]_q$ codes. The Griesmer defect of these codes is $1$. The weight distribution is in the following table.}\\

\begin{longtable}{|l|l|}
\caption{\label{tab:A-q-5-2} Weight distribution of optimal two-weight codes}       \\ \hline
Weight                               & Weight distribution            \\ \hline
$0$                                  & $1$                             \\ \hline
$(q-1)(q^{2m-1}-q^{m-1})$ & $2(q^m-1)$ \\ \hline
$(q-1)(q^{2m-1}-2q^{m-1})$ & $(q^m-1)^2$                        \\ \hline
\end{longtable}

{\bf Proof.} We only need to count hyperplanes in ${\bf F}_q^{2m}$ containing one $m$-dimensional subspace ${\bf F}_q^m \subset {\bf F}_q^{2m}$. This means that, the linear function defining $H$ is in the space generated by $m$ normal vectors of this space. Therefore, there are $2(q^m-1)$ such hyperplanes.\\

{\bf Example 5}  When $q=3$ and $m=2$, we construct an optimal two-weight $[64,4,42]_3$ codes, see \cite{Grassl}. This is a Griesmer defect one optimal code. Two weights are $42$ and $48$, $A_{42}=64$ and $A_{48}=16$. The optimal $[64,4,42]_3$ code documented in \cite{Grassl} is the shortening code of a linear $[65,5,42]_3$ code, which is from a stored generator matrix. Our code is simple and geometric.\\

Notice that the weight distribution of this infinite family of optimal two-weight $q$-ary linear codes is the same as the weight distribution Table II of optimal two-weight codes constructed in \cite[Theorem 6.1]{Shi}. Their construction via Gray images of two-weight codes over the finite ring ${\bf F}_p+u{\bf F}_p$, seems unnecessary, since their strong conditions on $q$ and $m$ are not essential. When $m=2r$, codes constructed in \cite[Proposition 5]{Hu} are these codes in Theorem 3.\\

Similarly, we have the following two result.\\

{\bf Corollary 3} {\em Let $m$ be a positive integer satisfying $m \geq 2$. Let $(k,{\bf u})=(2m+1,m,m+1)$. Then we construct an infinite family of optimal three-weight $[q^{2m+1}-q^{m+1}-q^m+1,2m+1, (q-1)(q^{2m}-q^m-q^{m-1})]_q$ codes. The weight distribution is in the following table.}\\

\begin{longtable}{|l|l|}
\caption{\label{tab:A-q-5-2} Weight distribution of optimal three-weight codes}       \\ \hline
Weight                               & Weight distribution            \\ \hline
$0$                                  & $1$                             \\ \hline
$(q-1)(q^{2m}-q^m)$ & $(q^{m+1}-1)$ \\ \hline
$(q-1)(q^{2m}-q^{m-1})$ & $(q^m-1)$ \\ \hline
$(q-1)(q^{2m}-q^m-q^{m-1})$ & $(q^{2m+1}-q^{m+1}-q^m+1$                        \\ \hline
\end{longtable}

{\bf Proof.} The Griesmer defect is one, from Theorem 1. We only need to prove that these codes are optimal. Otherwise, $$\Sigma_{i=0}^{2m} \lceil \frac{(q-1)(q^{2m}-q^m-q^{m-1})+1}{q^i} \rceil>\Sigma_{i=0}^{2m} \lceil \frac{(q-1)(q^{2m}-q^m-q^{m-1})}{q^i}\rceil+2,$$ since $2m\geq 4$. The conclusion is proved.\\

{\bf Example 6} When $q=3$ and $m=2$, we construct an optimal ternary three-weight $[208,5,138]_3$ code, see \cite{Grassl}. This is a Griesmer code. Three weights are $138,144, 156$, $A_{138}=208$, $A_{144}=26$ and $A_{156}=8$. The optimal $[208,5,138]_3$ code in \cite{Grassl} is from a stored generator matrix. \\

{\bf Theorem 4} {\em Let $m$ be a positive integer satisfying $m \geq 2$. Let $(k,{\bf u})=(3m,m,m,m)$. Then we construct an infinite family of optimal three-weight $[q^{3m}-3q^m+2,3m, (q-1)(q^{3m-1}-3q^{m-1})]_q$ codes. The Griesmer defect of these codes is $2$. The weight distribution is in the following table.}\\

\begin{longtable}{|l|l|}
\caption{\label{tab:A-q-5-2} Weight distribution of optimal three-weight codes}       \\ \hline
Weight                               & Weight distribution            \\ \hline
$0$                                  & $1$                             \\ \hline
$(q-1)(q^{3m-1}-q^{m-1})$ & $3(q^m-1)$ \\ \hline
$(q-1)(q^{3m-1}-2q^{m-1})$ & $3(q^{2m}-1)$   \\ \hline
$(q-1)(q^{3m-1}-3q^{m-1})$ & $q^{3m}-3q^{2m}-3q^m+5$   \\ \hline
\end{longtable}

{\bf Proof.} The Griesmer defect is upper bounded by $2$, from Theorem 2. The calculation of the weight distribution is the same as the proof of Theorem 3. For the optimality, we notice that the dimension of the code is at least three. Then $$\Sigma_{i=0}^{3m-1} \lceil \frac{(q-1)(q^{3m-1}-3q^{m-1})+1}{q^i}\rceil > q^{3m}-3q^m+2,$$ the conclusion is proved.\\

{\bf Example 7} In the case $m=1$, the code is near optimal and the weight distribution is the same. When $q=5$ and $m=1$, we construct an near optimal three-weight $[112,3,88]_5$ code.  Three weights are $88$, $92$ and $96$, $A_{88}=40$, $A_{92}=72$ and $A_{96}=12$. The optimal one in \cite{Grassl} is a linear $[112,3,90]_5$ code.\\

Similarly, we have the following result.\\

{\bf Theorem 5} {\em 1) Let $q>2$ be a prime power. Let $m$ be a positive integer. Let $(k,{\bf u})=(3m+1,m,m,m+1)$. Then we construct an infinite family of optimal five-weight $[q^{3m+1}-2q^m-q^{m+1}+2,3m+1, (q-1)(q^{3m}-2q^{m-1}-q^m)]_q$ codes. The weight distribution is in the following table. \\

2) In the case $q=2$, this infinite family of optimal binary codes have four weights, since the weight $q^{3m}-2q^{m-1}$ is the same in the weight $q^{3m}-q^m.$}\\

\begin{longtable}{|l|l|}
\caption{\label{tab:A-q-5-2} Weight distribution of optimal five-weight codes}       \\ \hline
Weight                               & Weight distribution            \\ \hline
$0$                                  & $1$                             \\ \hline
$(q-1)(q^{3m}-q^{m-1})$ & $2(q^m-1)$   \\ \hline
$(q-1)(q^{3m}-q^m)$ & $q^{m+1}-1$   \\ \hline
$(q-1)(q^{3m}-2q^{m-1})$ & $q^{2m}-1$   \\ \hline
$(q-1)(q^{3m}-q^{m-1}-q^m)$ &$2(q^{m+1}-q^m)$ \\ \hline
$(q-1)(q^{3m}-2q^{m-1}-q^m)$ & $q^{3m+1}-q^{2m}-3q^{m+1}+3$   \\ \hline
\end{longtable}

{\bf Example 8} When $q=2$ and $m=2$. Then an optimal $[114,7,56]_2$ code is constructed. The Griesmer defect is two. The four weights are $56,58,60,62$, $A_{56}=91$, $A_{58}=8$, $A_{60}=22$ and $A_{62}=6$. The optimal linear $[114,7,56]_2$ code documented in \cite{Grassl} is from a quasi-cyclic code. Our construction is structural.\\

Notice that this is not the binary Solomon-Stiffler codes invented in \cite{Solomon}, since the Griesmer defect is two. However, this binary Griesmer defect two code is optimal.\\

\section{Modified affine Solomon-Stiffler codes}

In affine Solomon-Stiffler codes constructed in Theorem 1, if a vector ${\bf v}$ is a column vector of the generator matrix $G$, then $\lambda {\bf v}$, where $\lambda \in {\bf F}_q^{*}$, is also a column vector of $G$. If we only pick up these columns of the form $\lambda {\bf v}$, where $\lambda$ takes over all elements of a subgroup of the order $e$ in the multiplicative group ${\bf F}_q^*$, we have the following result.\\

{\bf Theorem 6} {\em Let ${\bf C}$ be the code with the generator matrix $G$. Then ${\bf C}$ is a linear $[\frac{e(q^k-1-\Sigma_{i=1}^h (q^{u_i}-1))}{q-1}, k, e(q^{k-1}-\Sigma_{i=1}^h (q^{u_i-1})]_q$ code. The Griesmer defect of this code ${\bf C}$ is upper bounded similar to Theorem 2.}\\

{\bf Proof.} The proof is similar to the proof of Theorem 1 and Theorem 2.\\

The interesting point here is that Griesmer defects of these modified affine Solomon-Stiffler codes decrease. It is easy to see that when $e=1$ and $u_1<u_2<\cdots<u_h$, modified affine Solomon-Stiffler codes are $q$-ary projective Solomon-Stiffler codes invented in \cite{Solomon}.\\

{\bf Example 9} When $q=5$, $(k, {\bf u})=(4,1,1,1,1)$ and $e=1$. We construct a modified affine Solomon-Stiffler $[152, 4, 121]_5$ code. This is a linear Griesmer code. Notice that the punctured $[130,4, 99]_5$ code is obtained. The optimal linear $[130,4]_5$ code has the optimal distance $103$.\\

When $q=7$, $(k, {\bf u})=(3,1,1,1)$ and $e=1$. We construct a modified affine Solomon-Stiffler $[54,3,46]_7$ code. This is a linear Griesmer code. When $e=2$, we construct a linear $[108,3,92]_7$ code. This is also a linear Griesmer code. The punctured $[100, 3, 84]_7$ code is obtained. This code is almost optimal, see \cite{Grassl}.\\

When $q=8$, $(k, {\bf u})=(3,1,1,1)$ and $e=1$. We construct a modified affine Solomon-Stiffler $[70,3,61]_8$ code. This is a linear Griesmer code. When $e=2$, we construct a linear $[140,3,122]_7$ code. This is also a linear Griesmer code. The punctured $[130, 3, 112]_8$ code is obtained. This code is almost optimal, see \cite{Grassl}.\\

Let us consider the modified affine Solomon-Stiffler code associated with $(2m, m)$ and a divisor $e$ of $q-1$. This is a linear Griesmer $[\frac{e(q^{2m}-q^m)}{q-1}, 2m, e(q^{2m-1}-q^{m-1}]_q$ code with two nonzero weights $e(q^{2m-1}-q^{m-1})$ and $eq^{2m-1}$. The weight distribution can be calculated directly, $$A_{e(q^{2m-1}-q^{m-1})}=q^{2m}-q^m$$ and $$A_{eq^{2m-1}}=q^m-1.$$ Then the repetition two-copy of this code is a linear Griesmer $[\frac{2e(q^{2m}-q^m)}{q-1},2m, 2e(q^{2m-1}-q^{m-1})]_q$ code, with the weight distribution $$A_{2e(q^{2m-1}-q^{m-1})}=q^{2m}-q^m$$ and $$A_{2eq^{2m-1}}=q^m-1.$$ Notice that this is the same as the Griesmer codes constructed in \cite[Theorem 3.5]{Liu}. The weight distribution is the same as that in \cite[Theorem 3.3]{Liu}. The construction in \cite{Liu}, as that in \cite{Shi}, is technical, and some restrictions on $e$ and $m$ were needed.\\

{\bf Example 10} When $q=5$, $(k, {\bf u})=(4,1,1,1,1), e=1$, we construct a linear $[152,4,121]_5$ code. This code is Griesmer code and non-Solomon-Stiffler.\\

The optimal linear $[50,3,40]_5$ code documented in \cite{Grassl} is constructed from a constacyclic code. A projective Solomon-Stiffler code associated with $(k,u_1)=(3,2)$ has the same parameters.\\

In the following table, we list some non-Solomon-Stiffler codes constructed from Theorem 6. Most of them are optimal, see \cite{Grassl}. These codes have few nonzero weights, and their exact weight distributions can be determined explicitly.\\

\begin{longtable}{|l|l|l|l|}
\caption{\label{tab:A-q-5-3} Modified affine Solomon-Stiffler codes}\\ \hline
$q$&$[n,k,d]_q$&$(k,{\bf u})$&Optimality, comparing with \cite{Solomon}\\ \hline
$3$ &$[37,4,24]_3$& $(4,1,1,1)$ &optimal, non-Solomon-Stiffler\\ \hline
$3$ &$[116,5,76]_3$& $(5,1,1,1,1,1)$ &almost optimal, non-Solomon-Stiffler\\ \hline
$3$ &$[117,5,77]_3$& $(5,1,1,1,1)$ &almost optimal, non-Solomon-Stiffler\\ \hline
$3$ &$[118,5,78]_3$& $(5,1,1,1)$ &optimal, non-Solomon-Stiffler\\ \hline
$4$ &$[81,4,60]_4$& $(4,1,1,1,1)$ &optimal, non-Solomon-Stiffler\\ \hline
$4$ &$[83,4,62]_4$& $(4,1,1,1,1)$ &optimal, non-Solomon-Stiffler\\ \hline
$7$ &$[100,3,84]_7$& $(3,1,1,1)$ & almost optimal, non-Solomon-Stiffler \\ \hline
$8$ &$[130,3,112]_8$& $(3,1,1,1)$ & almost optimal, non-Solomon-Stiffler \\ \hline
\end{longtable}

\section{Few-weight Griesmer codes from the modified affine Solomon-Stiffler construction}

In this section, we construct some two-weight and three-weight Griesmer codes and determine their weight distributions.\\

{\bf Theorem 7} {\em Let $q>2$ be a prime power, $h$ and $m$ be a positive integers and $e$ be a divisor of $q-1$ satisfying $$e(h-1)<q.$$ Let $(k,{\bf u})=(hm, m, \ldots, m)$. Then we construct an infinite family of Griesmer $$[\frac{e(q^{hm}-hq^m+h-1)}{q-1},hm, e(q^{hm-1}-hq^{m-1})]_q$$ codes.}\\

{\bf Proof.} This follows immediately from Theorem 6.\\

{\bf Corollary 4} {\em Let $q$ be a prime power and $e$ be a divisor of $q-1$. Then we construct an infinite family of Griesmer $[\frac{e(q^{2m}-2q^m+1)}{q-1}, 2m, e(q^{2m-1}-2q^{m-1})]_q$ codes. The weight distribution is in the following table.}\\

\begin{longtable}{|l|l|}
\caption{\label{tab:A-q-5-2} Weight distribution of optimal two-weight codes}       \\ \hline
Weight                               & Weight distribution            \\ \hline
$0$                                  & $1$                             \\ \hline
$e(q^{2m-1}-q^{m-1})$ & $2(q^m-1)$ \\ \hline
$e(q^{2m-1}-2q^{m-1})$ & $(q^m-1)^2$                        \\ \hline
\end{longtable}

{\bf Proof.} This conclusion follows from Theorem 7 directly.\\

{\bf Corollary 5} {\em Let $m$ be a positive integer, $q>3$ be a prime power and $e$ be a divisor of $q-1$ satisfying $2e<q$. Then we construct an infinite family of Griesmer three-weight $[\frac{e(q^{3m}-3q^m+2}{q-1},3m, e(q^{3m-1}-3q^{m-1})]_q$ codes. The weight distribution is in the following table.}\\

\begin{longtable}{|l|l|}
\caption{\label{tab:A-q-5-2} Weight distribution of optimal three-weight codes}       \\ \hline
Weight                               & Weight distribution            \\ \hline
$0$                                  & $1$                             \\ \hline
$e(q^{3m-1}-q^{m-1})$ & $3(q^m-1)$ \\ \hline
$e(q^{3m-1}-2q^{m-1})$ & $3(q^{2m}-1)$   \\ \hline
$e(q^{3m-1}-3q^{m-1})$ & $q^{3m}-3q^{2m}-3q^m+5$   \\ \hline
\end{longtable}

{\bf Proof.} The conclusion follows from Theorem 4 and Theorem 7 immediately.\\

{\bf Example 11} When $q=9$, $(k, {\bf u})=(3,1,1,1)$ and $e=1$. We construct a three-weight Griesmer $[88,3,78]_9$ code. Three weights are $78$, $79$ and $80$, $A_{78}=464$, $A_{79}=240$ and $A_{80}=24$. This is a projective Solomon-Stiffler codes.\\

It is obvious that our geometric construction is more general and simple than ring-technique used in \cite{Shi,Liu}. Much more infinite families of Griesmer or optimal codes can be constructed. Weight distributions of these Griesmer or optimal codes can be determined geometrically.\\

\section{Modified affine Solomon-Stiffler codes when $u_1=u_2\cdots=u_h=1$, $h$ large}

In this section, we consider affine Solomon-Stiffler codes associated with $u_1=\cdots=u_h=1$, $h$ is a large positive integer satisfying $h<q^{k-1}$.  We prove the following result.\\

{\bf Theorem 8} {\em Let $q$ be a prime power and $h$ be a positive integer satisfying $k\leq h<q^{k-1}$. Suppose that $h$ lines span the whole space. If there exists a projective linear $[h,k]_q$ code with the maximum weight $h$. Then we can construct a linear $[q^k-hq+h-1,k, (q-1)(q^{k-1}-h)]_q$ code. If there exists no such a linear $[h,k]_q$ code with the maximum weight $h$, then the minimum weight of this code is at least $d \geq (q-1)(q^{k-1}-h+1)$.}\\

{\bf Proof.} From the proof of Theorem 1, the conclusion is equivalent to finding a nonzero vector vector ${\bf h}$ such that the inner products of ${\bf h}$ with these $h$ lines are nonzero. This is equivalent to the existence of a projective linear $[h,k]_q$ code with the maximum weight $h$.\\

{\bf Example 12} Let $q$ be a prime power satisfying $q>2k$. Let $h=k+\frac{k(k-1)}{2}$. These lines are spanned by vectors with one coordinate $1$ or two coordinates $1$. Then we can find a vector $(x_1, \ldots, x_k)$ satifying\\

1) $x_i \neq 0$;\\

2) $x_i+x_j\neq 0$, if $i \neq j$.\\

Then a linear $[q^k-hq+h-1,k, (q-1)(q^{k-1}-h)]_q$ code is constructed. The Griesmer defect is $(q-1)\lfloor \frac{h}{q} \rfloor$. When $k=4$ and $q=9$, $h=10$. A linear $[6480,4,5752]_9$ code is obtained. The Griesmer defect is $8$.\\

{\bf Corollary 6} {\em Let $q$ be a prime power and $h$ be a positive integer satisfying $k \leq h<q^{k-1}$. Suppose that $h$ lines span the whole space. If there exists a projective linear $[h,k]_q$ code with the maximum weight $h$. Then we can construct a projective linear $[\frac{q^k-hq+h-1}{q-1},k, (q^{k-1}-h)]_q$ code. The Griesmer defect of this code is upper bounded by $\Sigma_{i=0}^{k-1} \lfloor \frac{h}{q^i} \rfloor.$}\\

{\bf Proof.} The proof is similar to the argument of Theorem 8.\\

Though the projective linear codes described in Corollary 6 is not projective Solomon-Stiffler codes in \cite{Solomon}. When $h=2q$, the Griesmer defect is at most $2+\lfloor \frac{h}{q^2} \rfloor$. This linear code is almost optimal.\\

{\bf Corollary 7} {\em Let $q$ be a prime power and $h=2q$, and $k$ be a positive integer satisfying $k \geq 2q$. Suppose that these $2q$ lines are linear independent. Then we construct a projective linear almost optimal $[\frac{q^k-1}{q-1}-2q, k, d \geq q^{k-1}-2q]_q$ code.}\\

{\bf Proof.} From the proof of Theorem 1, we have $d\geq q^{k-1}-2q$. If the maximal possible distance of a linear $[\frac{q^k-1}{q-1}-2q, k]_q$ is $q^{k-1}-2q+2$, then from the Griesmer bound,  $\frac{q^k-1}{q-1}-2q \leq q^{k-1}-2q+2+q^{k-2}-1+q^{k-3}+\cdots+q+1=\frac{q^k-1}{q-1}-2q+1$. This is a contradiction.\\

When $q=2$, $h=4$, we construct a linear $[2^k-5, k, d \geq 2^{k-1}-4]_2$ code with the Griesmer defect at most $3$. This code is almost optimal, since $$\Sigma_{i=0}^{k-1}\lceil \frac{2^{k-1}-2}{2^i} \rceil=2^k-4>2^k-5.$$ For example, binary linear $[59,6,28]_2$ code is constructed. The optimal minimum weight of linear $[59,6]_2$ code is 29, see \cite{Grassl}.\\

When $q=3, h=6$, we construct an infinite family of linear $[\frac{3^k-1}{2}-6,k, 3^{k-1}-6]_3$ code with the Griesmer defect at most $2$. This code is almost optimal, since $$\Sigma_{i=0}^{k-1}\lceil \frac{3^{k-1}-4}{3^i} \rceil=\frac{3^k-1}{2}-5>\frac{3^k-1}{2}-6.$$ For example, binary linear $[34,4,21]_3$ code is constructed. The optimal minimum weight of linear $[34,4]_3$ code is 22, see \cite{Grassl}.\\

These almost optimal projective linear codes constructed in Corollary 7 are not included in \cite{Hu1}.

\section{Distance-optimal codes as codimension one or two subcodes in Griesmer codes}

We prove that any $k-1$ dimension subcodes of certain binary Griesmer $[n,k,2d]_2$ codes are distance-optimal (with respect to the Griesmer bound).\\

{\bf Theorem 9} {\em Let ${\bf C}$ be a binary linear Griesmer $[n,k,2d]_2$ code satisfying $k \geq 3$ and $d\leq 2^{k-2}$. Then any dimension $k-1$ subcode of ${\bf C}$ has minimum weight $2d$ and is distance-optimal. The Griesmer defect of this subcode is $1$.}\\

{\bf Proof.} Suppose there exists a linear $[n,k-1,2d+1]_2$ code. From the Griesmer bound, $$n \geq \Sigma_{i=0}^{k-2} \lceil \frac{2d+1}{2^i} \rceil \geq 2+\Sigma_{i=0}^{k-2} \lceil \frac{2d}{2^i} \rceil =1+\Sigma_{i=0}^{k-1} \lceil  \frac{2d}{2^i} \rceil.$$ Since $$n=\Sigma_{i=0}^{k-1} \lceil  \frac{2d}{2^i} \rceil,$$ this is a contradiction.\\

Consider binary Solomon-Stiffer code associated with $(k, u)$, $k>u\geq 3$, then we construct a linear Griesmer $[2^k-2^u,k,2^{k-1}-2^{u-1}]_2$ code, from Theorem 1. From Theorem 8, any $k-1$ dimensional subcode is a linear distance-optimal $[2^k-2^u, k-1, 2^{k-1}-2^{u-1}]_2$ code.\\

Distance-optimal codes constructed in \cite[Theorem 2]{Shi2} and \cite[Proposition 5.3, 5)]{ML} have the same parameters as above codes. Their constructions are complicated and technique. Actually, more general such distance-optimal binary codes are constructed in the following result. By the way, optimal binary three-weight $[3 \cdot 2^{3m-1},3m,3 \cdot 2^{3m-2}]_2$ codes constructed in \cite{LiShi} have the same parameters as repetition codes of the 1st order RM $[2^{3m-1},3m, 2^{3m-2}]_2$ codes.\\

{\bf Corollary 8} {\em Let ${\bf C}$ be a binary Solomon-Stiffler $[2^k-1-\Sigma_{i=1}^h (2^{u_i}-1),k, 2^{k-1}-\Sigma_{i=1}^h 2^{u_i-1}]_2$ code constructed in \cite{Solomon}. We assume that $k \geq 3$, $2 \leq u_1<u_2<\cdots <u_h<k$, Then we construct many dimension $k-1$ distance-optimal binary
$[2^k-1-\Sigma_{i=1}^h (2^{u_i}-1),k-1, 2^{k-1}-\Sigma_{i=1}^h 2^{u_i-1}]_2$ codes. The Griesmer defect is one.}\\

{\bf Proof.} The conclusion follows from Theorem 9 immediately, since binary Solomon-Stiffler codes are Griesmer.\\

Similarly we consider codimension two subcodes of a linear Griesmer code.\\

{\bf Theorem 10} {\em Let ${\bf C}$ be a binary linear Griesmer $[n,k,4d]_2$ code satisfying $k \geq 4$ and $d\leq 2^{k-4}$. Then any dimension $k-2$ subcode of ${\bf C}$ has minimum weight $4d$ and is distance-optimal. The Griesmer defect of $k-2$ dimension subcode is two.}\\

{\bf Proof.} If the codimension two subcode is not distance-optimal, then there exists a linear $[n,k-2,4d+1]_2$ code. From the Griesmer bound, $$n \geq \Sigma_{i=0}^{k-3} \lceil \frac{4d+1}{2^i} \rceil \geq 3+\Sigma_{i=0}^{k-3} \lceil \frac{4d}{2^i} \rceil =1+\Sigma_{i=0}^{k-1} \lceil \frac{4d}{2^i} \rceil.$$ Since $$n=\Sigma_{i=0}^{k-1} \lceil  \frac{4d}{2^i} \rceil,$$ this is a contradiction.\\

Hence there are many codimension two distance-optimal subcodes in binary Solomon-Stiffler codes.\\

{\bf Corollary 9} {\em Let ${\bf C}$ be a binary Solomon-Stiffler $[2^k-1-\Sigma_{i=1}^h (2^{u_i}-1),k, 2^{k-1}-\Sigma_{i=1}^h 2^{u_i-1}]_2$ code constructed in \cite{Solomon}. We assume that $k \geq 4$, $u_h=k-1$ and  $3 \leq u_1<u_2<\cdots <u_{h-1} \leq k-2$, Then we construct many dimension $k-2$ distance-optimal binary
$[2^k-1-\Sigma_{i=1}^h (2^{u_i}-1),k-1, 2^{k-1}-\Sigma_{i=1}^h 2^{u_i-1}]_2$ codes. The Griesmer defect of the dimension $k-2$ subcode is two.}\\

For $q$-ary linear Griesmer codes, similar results can be proved.\\

{\bf Theorem 11} {\em 1) Let ${\bf C}$ be a $q$-ary linear Griesmer $[n,k,qd]_q$ code satisfying $k \geq 3$ and $d\leq q^{k-2}$. Then any dimension $k-1$ subcode of ${\bf C}$ has minimum weight $qd$ and is distance-optimal. The Griesmer defect of this codimension one subcode is one.\\

2) Let ${\bf C}$ be a $q$-ary linear Griesmer $[n,k,q^2d]_q$ code satisfying $k \geq 4$ and $d\leq q^{k-4}$. Then any dimension $k-2$ subcode of ${\bf C}$ has minimum weight $q^2d$ and is distance-optimal. The Griesmer defect of this codimension two subcode is two}\\

From $q$-ary projective Solomon-Stiffler codes satisfying certain restrictions, many distance-optimal $q$-ary linear codes are constructed. All these distance-optimal codes have positive Griesmer defects.\\

Results in this section shows that, when some optimal codes with new parameters were constructed, it would be better to check if these distance-optimal codes (with respect to the Griesmer bound) are small codimension subcodes of known Griesmer codes.\\

\section{Distance-optimal codes as punctured codes of Griesmer codes}

In this section, we prove the following result for punctured codes of $q$-ary Griesmer codes.\\

{\bf Theorem 12} {\em 1) Let ${\bf C}$ be a $q$-ary linear Griesmer $[n,k,qd]_q$ code satisfying $d=qd'>1$. Then the punctured code at a position in the support of a minimum weight codeword of ${\bf C}$ is a linear $[n-1,k,qd'-1]_q$ Griesmer code.\\

2) Let ${\bf C}$ be a $q$-ary linear Griesmer $[n,k,qd'+1]_q$ code satisfying $k \geq 3$. Suppose that $d'$ satisfies that $d'\geq 1$, $d' \equiv d''$ $mod$ $q$, where $d''<q-1$.  Then the punctured code ${\bf C}$ at a position in the support of a minimum weight codeword of ${\bf C}$ has the Griesmer defect one. It is distance-optimal.}\\

{\bf Proof.} We have $$\Sigma_{i=0}^{k-1}\lceil \frac{qd'-1}{q^i} \rceil=qd'-1+\lceil \frac{qd'-1}{q}\rceil+\cdots+\lceil \frac{qd'-1}{q^{k-1}} \rceil=\Sigma_{i=0}^{k-1}\lceil \frac{qd'}{q^i}-1=n-1.$$ The conclusion 1) is proved.\\

Suppose that $d'=qd_1+d''$, where $d''<q-1$. Then, $$\Sigma_{i=0}^{k-1} \lceil \frac{qd'}{q^i} \rceil=qd'+d'+\lceil \frac{q^2d_1+qd''}{q^2}\rceil+\cdots+\lceil \frac{q^2d_1+qd''}{q^{k-1}} \rceil=\Sigma_{i=0}^{k-1} \lceil \frac{qd'+1}{q^i}\rceil-2.$$ The Griesmer defect of the punctured code is one.\\

If there is a linear $[n-1,k,qd'+1]_2$ code, then it is a contradiction to the condition that the linear $[n,k,qd'+1]_q$ code is Griesmer. Therefore the punctured $[n-1,k,qd']_q$ code is distance-optimal.\\

When Theorem 12 is applied to binary Solomon codes, then we have the following result.\\

{\bf Corollary 10} {\em Let ${\bf C}$ be a binary Solomon-Stiffler $[2^k-1-\Sigma_{i=1}^h (2^{u_i}-1),k, 2^{k-1}-\Sigma_{i=1}^h 2^{u_i-1}]_2$ code constructed in \cite{Solomon}. We assume that $k \geq 3$, $u_1=1,u_2=2$ and  $3 \leq u_3<\cdots <u_{h-1} \leq k-1$, Then we construct many dimension $k$ distance-optimal binary
$[2^k-1-\Sigma_{i=1}^h (2^{u_i}-1),k, 2^{k-1}-\Sigma_{i=1}^h 2^{u_i-1}-1]_2$ codes, as punctured codes. The Griesmer defect of these punctured codes is one.}\\

\section{Conclusion}

In this paper, we propose a general construction of Griesmer and optimal linear codes as an affine version of Solomon-Stiffler codes. Modified affine Solomon-Stiffler codes including projective Solomon-Stiffler codes as special cases, were given. Griesmer or optimal codes constructed in \cite{Shi,Liu} have the same parameters and weight distributions as very special case codes in our construction. Optimal binary codes constructed in a recent paper \cite{ML} and another paper \cite{Shi2} and codimension one subcodes of special binary Solomon-Stiffler codes have the same parameters. Distance-optimal codes as punctured or shortened codes of Griesmer codes were also discussed. We reconstructed many optimal codes in \cite{Grassl} from our modified affine Solomon-Stiffler codes. Projective linear almost optimal codes similar to codes in \cite{Solomon} were also constructed.\\

\end{document}